\documentclass[10pt,letterpaper]{article}

\usepackage{opex3}
\usepackage{times}
\usepackage{hyperref}
\usepackage{graphicx}
\usepackage{longtable}
\usepackage{color}
\usepackage{amssymb}
\usepackage{srcltx}
\newcommand{\ket}[1]{|#1\rangle}

\newcommand{\eq}{\begin{equation}}
\newcommand{\fine}{\end{equation}}

\newcommand{\aac}{\'a}

\begin{document}
\title{Generation of hybrid polarization-orbital angular momentum entangled states}

\author{Eleonora Nagali$^1$ and Fabio Sciarrino$^{1,2}$}

\address{$^1$Dipartimento di Fisica, Sapienza Universit\`{a} di Roma, Roma 00185, Italy}

\address{$^2$Istituto Nazionale di Ottica (INO-CNR), L.go E. Fermi 6, Florence 50125, Italy}

\email{fabio.sciarrino@uniroma1.it}

\begin{abstract}
Hybrid entangled states exhibit entanglement between different degrees of freedom of a particle pair and thus could be useful for asymmetric optical quantum network where the  communication channels are characterized by different properties.
We report the first experimental realization of hybrid polarization-orbital angular momentum (OAM) entangled states by adopting a spontaneous parametric down conversion source of polarization entangled states and a polarization-OAM transferrer. The generated quantum states have been characterized through quantum state tomography. Finally, the violation of Bell's inequalities with the hybrid two photon system has been observed.
\end{abstract}


\section{Introduction}
The development of tailored photonic sources suitable to produce entanglement  represents a crucial resource for quantum information applications like quantum communication schemes, quantum cryptographic protocols, and for fundamental tests of quantum theory \cite{Gisi02}. Parametric down conversion has been proven to be the best source of entangled photon pairs so far in an ever increasing number of experiments on the foundations of quantum mechanics and in the new field of quantum
communication \cite{DeMa05}. Optical implementation of quantum information processing have been realized by several, different approaches, each one with its own advantages and limitations concerning the generation, manipulation, transmission, detection, robustness of the information carriers. While initially most of the effort has been devoted to the implementation of polarization entangled states \cite{Kwia95,Kwia99,Cine04}, in the last few years entangling different degrees of freedom has attracted much attention. Within this scenario, the orbital angular momentum (OAM), the degree of freedom of light associated with rotationally structured transverse spatial modes, has been recently exploited 
to encode quantum states \cite{Moli07,Fran08,Barr08,Naga09c}. Generation of OAM-entangled pairs of photons has been demonstrated mainly by spontaneous parametric down-conversion \cite{Mair01,Vazi03,Torr03,Moli04,Lang04,Moli05,Barr05,Oemr05,Lany08}. By merging different techniques, it is possible to exploit the power and the advantages of each method and hence overcome the present technological limitations. 

Hybrid entangled states exhibit entanglement between different degrees of freedom of a particle pair. The generation of such states can be useful for asymmetric optical quantum network where the different communication channels adopted for transmitting quantum information exhibit different properties. In such a way one could adopt the suitable degree of freedom with larger robustness  along the channel. From a fundamental point of view, the observation of non-locality with hybrid systems proves the fundamental independence of entanglement from the physical realization of the adopted Hilbert space. Very recently the hybrid entanglement of photon pairs between the path (linear momentum) of one photon and the polarization of the other photon has been reported by two different techniques \cite{Ma09,Neve09}. Nevertheless, the capability of generating hybrid-entangled state encoded in the polarization and OAM of single photons could be advantageous since it could allow the engineering of qubit-qudit entangled states, related to the different Hilbert space dimensionality of the two degrees of freedom. It has been pointed out that such states are desiderable for quantum information and communication protocols, as quantum teleportation, and for the possibility to send quantum information through an optical quantum network composed by optical fiber channels and free-space \cite{Neve09,Chen09}.

\begin{figure}[t]
\centering
\includegraphics[scale=.3]{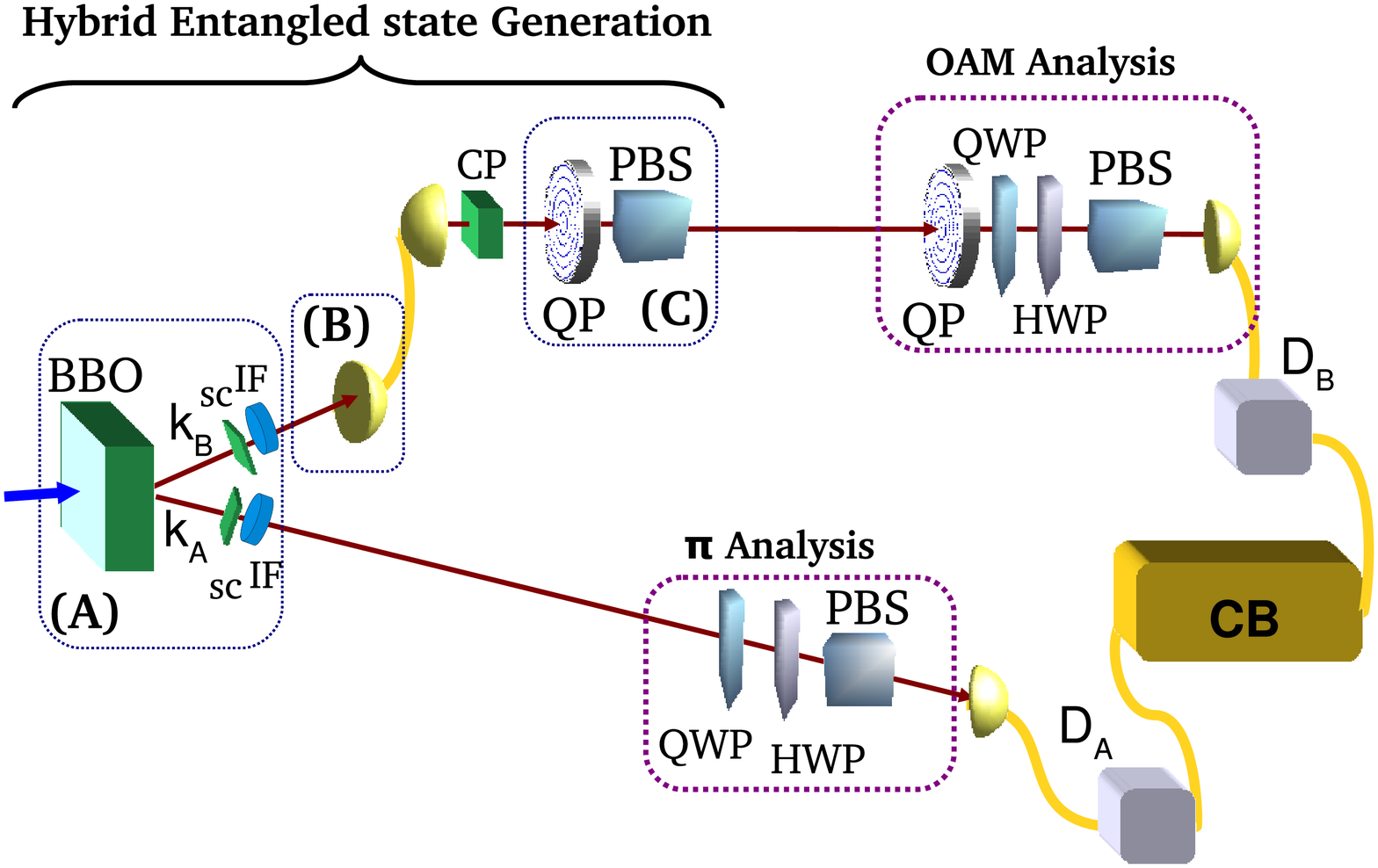}
\caption{Experimental setup adopted for the generation and characterization of hybrid $\pi$-OAM entangled states.\textbf{(A)}Generation of polarization entangled photons on modes $k_A$ and $k_B$.\textbf{(B)}Projection on the OAM state with $m=0$ through the coupling on a single mode fiber (SMF).\textbf{(C)}Encoding of the state in the OAM subspace $o_2$ through the $\pi\rightarrow o_2$ transferrer.}
\label{setup}
\end{figure}

In this paper, we report the experimental realization of hybrid polarization-OAM entangled states, by adopting the deterministic polarization-OAM transferrer introduced in Ref.\cite{Naga09a,Naga09b}. Polarization entangled photon pairs are created by spontaneous parametric down conversion, the spatial profile of the twin photons is filtered through single mode fibers and finally the polarization is coherently transferred  to OAM state for one photon.  A complete characterization of the hybrid entangled quantum states has been carried out by adopting the quantum state tomography technique. This result, together with the achieved generation rate, the easiness of alignment and the high quality of the generated state, can make this optical source a powerful tool for advanced quantum information tasks.
For instance the OAM features can be more appropriate for mapping single photon states in atomic systems. 

\section{Experimental apparaturs and generation of hybrid states}

Let us now describe the experimental layout shown in Fig.\ref{setup}. A $1.5 mm$ thick $\beta$-barium borate
crystal (BBO) cut for type-II phase matching \cite{Kwia95}, is pumped by the 
second harmonic of a Ti:Sa mode-locked laser beam, and generates via
spontaneous parametric fluorescence polarization entangled photon pairs on modes $k_A$ and $k_B$ with wavelength $\lambda=795$ nm, and pulse bandwidth $\Delta\lambda=4.5$ nm, as determined by two interference filters (IF). The
spatial and temporal walk-off is compensated by inserting a $\frac{\lambda }{%
2}$ waveplate and a $0.75$ mm thick BBO crystal (SC) on each output mode $k_{A}$
and $k_{B}$ \cite{Kwia95}. Thus the source generates photon pair in the singlet entangled state encoded in the polarization, i.e. $\frac{1}{\sqrt{2}}(\ket{H}^{A}\ket{V}^{B} -\ket{V}^{A}\ket{H}^{B})$.
The photon generated on mode $k_A$ is sent through a standard polarization analysis setup and then coupled to a single mode fiber connected to the single-photon counter modules (SPCM) $D_A$. The photon generated on mode $k_B$ is coupled to a single mode fiber, in order to collapse its transverse spatial mode into a pure TEM$_{00}$, corresponding to OAM $m=0$. After the fiber output, two waveplates compensate (CP) the polarization rotation introduced by the fiber. To transform the polarization entangled pairs into an hybrid entangled state the photon $B$ is sent through the quantum transferrer $\pi\rightarrow o_2$, which transfers the polarization quantum states in the OAM degree of freedom.
 \begin{figure}[t!!]
\centering
\includegraphics[scale=.3]{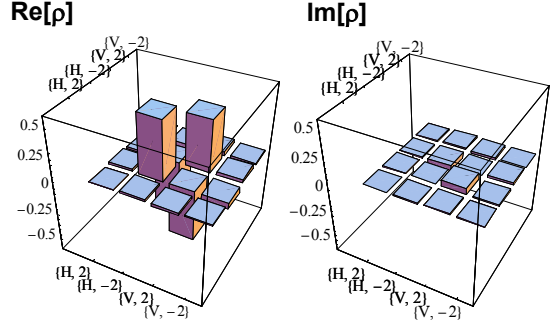}
\caption{Experimental density matrix of the  hybrid entangled state generated after the transferrer transformation on photons on $k_B$ mode. Each measurement setting lasted $15$s.}
\label{tomo}
\end{figure}
The quantum transferrers have been extensively described in \cite{Naga09a,Naga09b}. To sum up, the transformation $\ket{\varphi}_{\pi}\ket{0}_o\rightarrow\ket{H}_{\pi}\ket{\varphi}_{o_2}$ carried out by the transferrer, is achieved through a q-plate device,  which couples the spinorial (polarization) and orbital contributions of the angular momentum of photons\cite{Marr06,Naga09a}.  
%A q-plate (QP) is a birefringent slab having a suitably patterned
%transverse optical axis, with a topological singularity at its center \cite{Marr06}. The ``charge'' of this singularity reads $q=1$, which is determined by the
%(fixed) pattern of the optical axis. For a birefringent retardation $\delta_{QP}=\pi$ uniform across the device, a QP modifies the OAM state $m$ of a light beam
%crossing it, imposing a variation $\Delta{m}={\pm}2$ whose sign
%depends on the input polarization, positive for left-circular and
%negative for right-circular. On the polarization degree of freedom, the q-plate acts as a half-waveplate. Hence, an input TEM$_{00}$ mode (having $m=0$) is converted into a beam with $m=\pm2$.
Here and after, we will denote the bidimensional OAM subspace with $m=\pm 2$, where $m$ denotes here the OAM per photon along the beam axis in units of $\hbar$,  as $o_2=\{\ket{+2},\ket{-2}\}$.
According to the nomenclature $\ket{\varphi}_{\pi}\ket{\phi}_{o_2}$, the $\ket{\cdot}_{\pi}$ and $\ket{\cdot}_{o_2}$ stand for the photon quantum state `kets' in the polarization and OAM degrees of freedom.  Following the same convention, the OAM equivalent of the two basis linear polarizations $\ket{H}$ and $\ket{V}$ are then
defined as $\ket{h}=(2^{-1/2})(\ket{+2}+\ket{-2})\nonumber ; \ket{v}=(2^{-1/2})(\ket{+2}-\ket{-2})$.
Finally, the $\pm45^{\circ}$ angle ``anti-diagonal'' and
``diagonal'' linear polarizations will be hereafter denoted with the
kets $\ket{+}=(2^{-1/2})(\ket{H}+\ket{V})$ and
$\ket{-}=(2^{-1/2})(\ket{H}-\ket{V})$, and the corresponding OAM
states are defined analogously:
$\ket{a}=e^{-i\pi/4}(\ket{+2}+i\ket{-2})(2^{-1/2})\nonumber; \ket{d}=e^{i\pi/4}(\ket{+2}-i\ket{-2})(2^{-1/2})$.
The transformation established by a q-plate with $q=1$, as the one adopted in our experiment, can be described as:
\begin{eqnarray}
\ket{L}_{\pi}\ket{0}_{o} & \stackrel{QP}{\rightarrow} & \ket{R}_{\pi}\ket{+2}_{o_2} \nonumber\\
\ket{R}_{\pi}\ket{0}_{o} & \stackrel{QP}{\rightarrow} &
\ket{L}_{\pi}\ket{-2}_{o_2} \label{eqqplate}
\end{eqnarray}
where $L$ and $R$ denote the left and right circular polarization states, respectively. Any coherent superposition of the two input states given in Eq.(\ref{eqqplate}) is preserved by the QP transformation, leading to the equivalent superposition of the
corresponding output states \cite{Naga09a}.
Thus by combining the transformation induced by the q-plate and a polarizing beamsplitter the map $\ket{\varphi}_{\pi}\ket{0}_{o} \rightarrow \ket{H}_{\pi}\ket{\varphi}_{o_2}$ can be achieved with an efficiency of conversion equal to $50\%$. It is possible to realize a fully \emph{deterministic} transferrer $\pi\rightarrow o_2$ at the price of a more complex optical layout, based on a q-plate and a
Mach-Zehnder interferometer, as shown in \cite{Naga09b}. 
After the transferrer operation the polarization entangled state is transformed into the hybrid entangled state:
\begin{equation}
\frac{1}{\sqrt{2}}(\ket{H}^{A}_{\pi}\ket{+2}^{B}_{o_2} -\ket{V}^{A}_{\pi}\ket{-2}^{B}_{o_2})\ket{0}^{A}_{o}\ket{H}^{B}_{\pi}
\label{hybridstate}
\end{equation}
 In order to analyze with high efficiency the OAM degree of freedom, we exploited the $o_2\rightarrow\pi$ transferrer, as shown in \cite{Naga09b,Naga09c}. By this approach any measurement on the OAM state is achieved by measuring the polarization after the transferrer
device, as shown in Fig.\ref{setup}. Finally the photon has been coupled to a single mode fiber and then detected by $D_B$ connected to the coincidence box (CB), which records the coincidence counts between $[D_A,D_B]$. We observed a final coincidence rate equal to $C=100 coinc/s$ within a coincidence window of 3 ns. This experimental data is in agreement with the expected value, determined from $C_{source}=6 kHz$ after taking
into account two main loss factors: hybrid state preparation probability
$p_{prep}$, and detection probability $p_{det}$. $p_{prep}$ depends on the conversion
efficiency of the q-plate ($0.80\pm0.05$) and on the probabilistic
efficiency of the quantum transferrer $\pi\rightarrow o_{2}$ $(0.5)$,
thus leading to $p_{prep}=0.40\pm0.03$. The detection efficiency includes the q-plate conversion efficiency $(0.8)$, the transferrer  $o_{2}\rightarrow\pi$ $(0.5)$, and the  single mode fiber coupling $(0.2)$. Hence $p_{det}=0.08$. The observed experimental rate includes a reduction factor $\sim 8$ due to the adoption of probabilistic transferrers \cite{Naga09b}, and by achieving a higher single mode fiber coupling efficiency. Hence, by modifying the transferrers, we expect to achieve a detection rate equal to about $800 coinc/s$.

\section{Characterization of the state}
To completely characterize the state in Eq. \ref{hybridstate} we reconstructed the density matrix of the quantum state. The tomography reconstruction requires the estimation of 16 operators \cite{Jame01} through 36 separable measurements on the polarization-OAM subspaces. 
We carried out the reconstruction of the density matrix $\rho_{\pi,o_2}^{A,B}$ after the polarization-OAM conversion. The experimental results are reported in Fig.\ref{tomo}, with the elements of the density matrices expressed in the polarization and OAM basis
$\{\ket{H,+2},\ket{H,-2},\ket{V,+2},\ket{V,-2}\}$. 
The fidelity with the singlet states $\ket{\Psi^{-}}$ has been evaluated to be
$F(\ket{\Psi^{-}},\rho_{\pi,o_2}^{A,B})=(0.957\pm 0.009)$, while the experimental linear entropy of the state reads $S_L=(0.012\pm0.002)$. A more quantitative parameter associated to the generated polarization-entangled states is given
by the concurrence $C=(0.957\pm 0.002)$. These values demonstrate the high degree of hybrid entanglement generation.

\begin{figure}[t!]
\centering
\includegraphics[scale=.32]{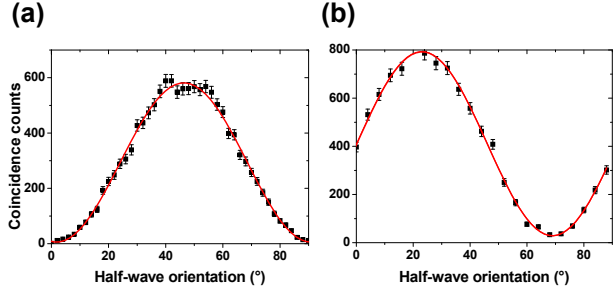}
\caption{Coincidence rate $[D_A,D_B]$ measured as a function of the angle $\theta$ of the half wave plate on the arm $k_A$ for OAM detected state \textbf{(a)} $\ket{+2}$ and \textbf{(b)} $\ket{h}_{o_2}$.}
\label{bell}
\end{figure}

To further characterize the hybrid quantum states, the violation of Bell's inequalities with the two photon system have been addressed.
First, we measured the photon coincidence rate as a function of the orientation of the half-wave plate on Alice arm for two different OAM basis analysis, namely  $\{\ket{+2}_{o_2},\ket{-2}_{o_2}\}$ and $\{\ket{h}_{o_2},\ket{v}_{o_2}\}$. The variation of the number of coincidences $N(\theta)$ with the
angle $\theta$ is in agreement with the one expected for entangled states
such as $N(\theta)=N_0(1+cos\theta)$: Fig.\ref{bell}. The coincidence fringe visibility reaches the values $V =(0.966\pm 0.001)$ and $V =(0.930\pm 0.007)$.
Hence, a non-locality test, the CHSH one \cite{Clau69}, has been carried out. Each of two partners, A (Alice) and B (Bob) measures a dichotomic observable among two possible ones, i.e. Alice randomly measures either \textbf{a} or \textbf{a'} while Bob measures \textbf{b} or \textbf{b'}, where the outcomes of each measurement are either $+1$ or $-1$. For any couple of measured observables $(A = \{\textbf{a}, \textbf{a'}\}, B =\{\textbf{b}, \textbf{b'}\})$, we define the following correlation function
$E(A,B) = \frac{N(+,+) + N(-,-) - N(+,-) - N(-,+)}{N(+,+) + N(-,-) + N(+,-) + N(-,+)}$
where $N(i,j) $ stands for the number of events in which the observables $A$ and $B$ have been found equal to the
dichotomic outcomes $i$ and $j$. Finally we define the parameter $S$ which takes into account the correlations for the
different observables 
\begin{equation}
S = E(\textbf{a}, \textbf{b}) + E(\textbf{a'}, \textbf{b}) + E(\textbf{a}, \textbf{b'}) - E(\textbf{a'}, \textbf{b'})
\end{equation}
Assuming a local realistic theory, the relation $|S|\leq S_{CHSH}=2$ holds.
To carry out a non-locality test in the hybrid regime, we define the two sets of dichotomic observables
for A and B. For Alice the basis \textbf{a} and \textbf{a'} correspond, respectively, to the linear polarization basis $\{\ket{H}_{\pi},\ket{V}_{\pi}\}$ and $\{\ket{+}_{\pi},\ket{-}_{\pi}\}$. For Bob the basis \textbf{b} and \textbf{b'} correspond, respectively, to the OAM basis $\{cos(\frac{\pi}{8})\ket{+2}-sin(\frac{\pi}{8})\ket{-2},-sin(\frac{\pi}{8})\ket{+2}+cos(\frac{\pi}{8})\ket{-2}\}$ and $\{cos(\frac{\pi}{8})\ket{+2}+sin(\frac{\pi}{8})\ket{-2},sin(\frac{\pi}{8})\ket{+2}-cos(\frac{\pi}{8})\ket{-2}\}$.
Experimentally we obtained the following value by
carrying out a measurement with a duration of $60 s$
and an average statistics per setting equal to about $1500$ events:
$S =(2.51\pm0.02)$. Hence a violation by more than 25 standard deviation over
the value $S_{CHSH} = 2$ is obtained. This experimental
value is in good agreement with an experimental visibility of $V=(0.930\pm0.007)$ which should lead to $S =(2.57\pm0.02)$.

\section{Conclusion}
In conclusion, we presented a source of polarization-OAM hybrid entanglement
based on SPDC source and $\pi\rightarrow o_2$ transferrer. We have shown that this system provides quantum states with high fidelity and with a bright generation rate. Moreover by adopting several concatenated q-plates the generation of hybrid states with higher OAM value could be obtained. Hybrid entangled states could be adopted to carry out quantum state teleportation between different degrees of freedom of light. Furthermore by inserting a quantum transferrer $\pi\rightarrow o_2$ also on mode $k_A$, a controllable source of OAM entangled states could be achieved.

This work was supported by FARI project, Finanziamento Ateneo 2009 of Sapienza Universit{\aac} di Roma, project PHORBITECH of the Future and Emerging Technologies (FET) programme within the Seventh Framework Programme for Research of the European Commission, under FET-Open grant number 255914, and project HYTEQ FIRB-"Futuro in Ricerca" (MIUR).

\end{document}